%Paper: hep-ph/9502206
%From: ellwange@qcd.th.u-psud.fr (Ellwanger)
%Date: Wed, 1 Feb 1995 14:20:22 GMT

\magnification=1200
\headline{\ifnum\pageno=1 \nopagenumbers
\else \hss\number \pageno \fi\hss}
\footline={\hfil}
\parindent=12pt
\baselineskip=20pt
\hsize=15 truecm
\vsize=23 truecm
\hoffset=0.7 truecm
\voffset=1 truecm
\overfullrule=0pt
%@@@@@@@@@@@@@@@@@@@@@@@@@@@@@@@@@@@@@@@@@@@@@@@
\noindent{January 1995}
%\rightline
\hfill{LPTHE Orsay 95-04}\par
\rightline{LPT Strasbourg 95-01}\par
\rightline {SPhT Saclay T95/04}\par
\vskip .5 truecm
\centerline{\bf HIGGS PHENOMENOLOGY OF THE SUPERSYMMETRIC}\par
\centerline{\bf MODEL WITH A GAUGE SINGLET} \par
\vskip 1 truecm
\centerline{\bf Ulrich ELLWANGER} \par
\centerline{Laboratoire de Physique Th\'eorique et Hautes
Energies\footnote{*}{Laboratoire
associ\'e au Centre National de la Recherche Scientifique - URA 63}} \par
\centerline{Universit\'e de Paris XI, b\^atiment 211, 91405 Orsay Cedex,
France} \par
\vskip .5 truecm
\centerline{\bf Michel RAUSCH DE TRAUBENBERG} \par
\centerline{Laboratoire de Physique Th\'eorique, Universit\'e Louis
Pasteur}\par
\centerline{ 3-5 rue de l'universit\'e, 67084 Strasbourg
Cedex, France} \par
\vskip .5 truecm
\centerline{\bf Carlos A. SAVOY} \par
\centerline{CEA, Service de Physique Th\'eorique, CE-Saclay, 91191
Gif-sur-Yvette Cedex, France} \par

\vskip 2 truecm
\noindent \underbar{{\bf Abstract}} \par
We discuss the Higgs sector of the supersymmetric standard model
extended by a gauge singlet for the
range of parameters, which is compatible with universal soft
supersymmetry breaking terms at the GUT
scale. We present results for the masses, couplings and decay properties
of the lightest Higgs
bosons, in particular with regard to Higgs boson searches at LEP. The
prospects differ
significantly from the ones within the MSSM.
 \par

%\vskip 1 truecm
%\noindent LPTHE Orsay 95-04 \par
%\noindent LPT Strasbourg 95-01\par
%\noindent SPhT Saclay T95/04\par
%\noindent January 1995
\vfill \supereject
\noindent 1. Introduction \par

The search for the Higgs boson belongs to the most interesting tasks of
future experiments such as
LEP 200. The prospects are particularly attractive in supersymmetric
models, where the lightest
neutral Higgs scalar cannot be too heavy. Most of the analysis of a
supersymmetric Higgs sector,
e.g. at LEP 1 [1], is performed within the minimal supersymmetric
standard model (MSSM). The Higgs
sector of the MSSM involves only two unknown parameters, which allows to
obtain relations between
the masses and couplings of the different particles [2]. (These
relations get somewhat modified,
however, due to radiative corrections which depend on additional
parameters as the softly
supersymmetry breaking interactions.) \par

In this paper we consider a modest extension of the MSSM, which amounts
to the addition of a gauge
singlet superfield to the Higgs sector [3-5]. Subsequently this model
will be called the (M+1)SSM. The (M+1)SSM has some
attractive theoretical features: the
superpotential can be
chosen to be scale invariant, thus there is no ``$\mu$-problem'' as in
the MSSM. Assuming
relations among the susy breaking terms at a large scale $M_{GUT}$ (such
as, e.g., universal
gaugino masses, scalar masses and trilinear scalar couplings) the model
has the same number of
free parameters as the MSSM in spite of the presence of the additional
singlet superfield. \par

It is evident, that in the (M+1)SSM the parameters in the Higgs
sector such as masses and
couplings to the $Z$-boson differ significantly from the MSSM. It is
thus desirable to interpret
the experimental findings independently from the relations between the
parameters within the
MSSM. On the other hand it would be helpful to have an idea of the
ranges of the masses and
couplings, which are theoretically allowed within the (M+1)SSM. \par

If one allows for arbitrary independent variations of all parameters of
the (M+1)SSM at the
weak scale, a large range particle masses and couplings can be obtained
[5-8]. There are obvious
constraints on the parameters, however, which should be imposed: the
effective potential,
e.g., has to have the correct properties: the minimum where the $SU(2)
\times U(1)$ symmetry is
broken as desired has to be the absolute minimum; charged and/or
coloured fields as sleptons,
squarks and charged Higgs scalars are not allowed to obtain vevs. In
addition present
experimental lower limits on sparticle masses should be satisfied. \par

Finally one can invoke theoretical prejudices such as universal gaugino
masses, scalar masses
and trilinear scalar couplings at $M_{GUT}$. A complete scan of the
pa\-ra\-me\-ter space of the (M+1)SSM,
which is consistent with all these constraints, has been
performed [9]. Recently also
certain deviations from universal susy breaking terms at $M_{GUT}$ have
been investigated [10],
but the corresponding sets of low energy parameters did not exceed the
ranges covered by the
assumption of universality. \par

In the present paper we will present results for the range of low energy
pa\-ra\-me\-ters within the
Higgs sector of the (M+1)SSM, which is obtained from the scan over
universal susy breaking terms
at $M_{GUT}$. We will focus on the masses of the lightest Higgs scalars
and pseudoscalars, their
couplings to the $Z$ boson, and comment on their decay properties. In
particular we will be
interested in the question which region of the parameter space is
accessible to LEP 2. \par

\noindent 2. The Model\par

The particle content of the Higgs sector is
given by the two MSSM Higgs doublet superfields $H_1$ and $H_2$, and the
additional gauge singlet
superfield $S$. The top quark sector is important because of its
radiative corrections to the
parameters of the Higgs sector. It involves the right handed top quark
$T_R$, and the left handed
doublet $Q$ containing the left handed top and bottom quarks $T_L$ and
$B_L$. The relevant part of
the superpotential is of the form
$$W = h_t \ Q \ H_2 \ T_R^c + \lambda \ H_1 \ H_2 \ S + {\kappa \over 3}
S^3 \eqno(1)$$
\noindent and involves three dimensionless Yukawa couplings $h_t$,
$\lambda$ and $\kappa$, but no mass
term. The only dimensionful parameters of the model are the
supersymmetry breaking gaugino masses,
scalar masses and trilinear couplings:
$$\left ( \mu_1 \lambda_1\lambda_1 + \mu_2 \lambda_2 \lambda_2 + \mu_3
\lambda_3 \lambda_3 +
h_t \ A_t \ Q \ H_2 \ T_R^c + \lambda A_{\lambda} \ H_1 \ H_2 \ S +
{\kappa \over 3} S^3 \right ) +
h.c.$$ $$+ m_1^2 |H_1|^2 + m_2^2 |H_2|^2 + m_S^2 |S|^2 + m_Q^2 |Q|^2 +
m_T^2 |T|^2 \eqno(2)$$
\noindent where $\lambda_1$, $\lambda_2$ and $\lambda_3$ are the
gauginos of the $U(1)_Y$,
$SU(2)$ and $SU(3)$ gauge groups, respectively. \par

The scalar potential contains the standard $F$ and $D$ terms, the
supersymmetry breaking terms
and in addition one loop radiative corrections of the form
$$V_{rad} = {1 \over 64 \pi^2} S \ tr \left [ M^4 \ell n \left ( M^2/Q^2
\right ) \right ]
\eqno(3)$$
\noindent where we only take the top quark and squark loops into
account. We include, however, the numerically important contributions
beyond the leading log approximation, which depend on $A_t$, the vev
of $S$ and the difference between $M_Q^2$ and $M_T^2$ [11]. $Q^2$
denotes the renormalization point of ${\cal O}(M_{Susy}^2)$.
\par

After minimization of the potential and the removal of the Goldstone
modes the physical
particle content in the Higgs sector is given by three neutral scalars,
two neutral
pseudoscalars and one charged Higgs boson. The corresponding mass
matrices in terms of the
parameters of the low energy effective potential can be found in [5-9].
In addition there are
two (Dirac-) charginos, and five two-component neutral fermionic states
(``neutralinos''). \par

As mentioned above, in this paper we constrain the range of the low
energy parameters of the
model by requiring universal supersymmetry breaking terms at $M_{GUT}
\sim 10^{16}$ GeV. Thus
we start by scanning over $\sim 10^6$ points in the five dimensional
parameter space of the
model at $M_{GUT}$, given by the three Yukawa couplings $\lambda_0$,
$\kappa_0$ and $h_{t0}$ and the
ratios of the supersymmetry breaking terms $m_0/\mu_0$, $A_0/\mu_0$. In
each case we integrate the
renormalization group equations down to the electroweak scale of ${\cal
O}(100)$ GeV, determining
thereby the parameters of the low energy theory appearing in eqs. (1)
and (2). \par

Next we minimize the low energy effective potential numerically in each
case, including the
radiative corrections eq. (3). We check, whether we have found the
absolute minimum of the
potential, and verify, whether squarks or sleptons do not assume vevs,
which would break color
and/or electromagnetism [4, 12]. \par

In the remaining cases we determine the overall scale of the
dimensionful parameters by identifying
$<H_1>^2 + <H_2>^2$ with $2 M_Z^2/(g_1^2 + g_2^2)$, and compute the
physical masses of all
particles. Then we impose the following experimental constraints:
concerning the top quark, we
require the pole mass $m_{top}$ to be just roughly two standard
deviations within the CDF value
[13], i.e. 150 GeV $< m_{top} <$ 200 GeV. We demand the charginos and
sneutrinos to be heavier than
45 GeV, and the neutralinos to be either heavier than 45 GeV or to
couple sufficiently weakly to
the $Z$ boson such that they do not contribute more than 7 MeV to its
invisible width. The lightest
neutralino is always the lightest sparticle within the range of
parameters obtained finally, and
the other sparticles turn out to be automatically sufficiently heavy
such that they satisfy the
present experimental limits. (In particular the charged Higgs boson is
heavier than 135 GeV, hence
it will play no role at LEP 2.) \par

\noindent 3. Higgs Masses and Couplings \par

The first question concerns the upper limit on the
mass of the lightest Higgs
boson $h$. Due to the radiative corrections to the scalar potential, eq.
(3), this upper limit
depends on the scale of the supersymmetry breaking terms, notably on the
stop masses $m_Q^2$,
$m_T^2$ and the trilinear coupling $A_t$. Within the present procedure,
however, no automatic
upper limit on the scale of supersymmetry breaking is obtained. On the
other hand the present
procedure makes it evident, that more and more fine tuning is required
for large scales of
supersymmetry breaking, i.e. the density of points in the parameter
space decreases in the
multi-TeV region. \par

This is visible from fig. 1, where we plot the mass of the lightest
neutral Higgs scalar $h$ versus
the mass of the gluino as a representative of the scale of supersymmetry
breaking, for $\sim $ 5
000 points in parameter space. One finds an upper limit on the mass of
the lightest Higgs boson
of $\sim$ 140 GeV for a gluino mass below $\sim$ 1 TeV, and an upper
limit of $\sim$ 160 GeV
for a gluino mass up to $\sim$ 2,5 TeV. (The upper limit on the Higgs
mass would decrease by
$\sim$ 10 GeV, if the top quark mass would be required to be lighter
than or equal to 175 GeV).
 From fig. 1 one finds that for a large part of the parameter space $h$
will be too heavy to be
produced at LEP 2. \par

On the other hand fig. 1 shows that for some part of the parameter space
the lightest Higgs
scalar can indeed be very light. (Note that we have not yet included
experimental constraints
from the unsuccessful Higgs search at LEP 1 at this stage). At this
point the question
arises, how easily a light Higgs boson can be seen in $Z$ boson decays
in this model. A
priori two possible processes exist:
$$ \eqalignno{
&a) \ Z \to Z + h  \cr
&b) \ Z \to A + h  &(4)\cr
}$$
\noindent where $A$ denotes a neutral pseudoscalar boson; in the
following $A$ will be the lightest one among the two
pseudoscalars, which exist in the (M+1)SSM. (The other one turns out
to be heavier than 120 GeV and will accordingly play no role at
LEP2.)
Let us first have a look at process a). Generally
the lightest Higgs boson $h$ is a superposition of three neutral scalar
fields: $$h = c_1 \ h_1 + c_2
\ h_2 + c_3 \ s \ \ \ . \eqno(5)$$

\noindent Only the $SU(2)$ doublets $h_1$ and $h_2$ (with hypercharges
$\pm$ 1/2) couple to
the $Z$ boson, the singlet $s$ has no gauge boson couplings. It has been
noted before [5,
6], that the lightest Higgs boson could be dominantly a gauge singlet in
the (M+1)SSM.
In fig. 2 we show a plot of the coefficient $c_3$ versus the mass of $h$
for the present
sample of points in parameter space. We observe, that indeed the
parameter space can be
approximately divided into two distinct regions: a region, where $h$ is
dominantly gauge
singlet ($c_3$ is close to 1) and possibly very light, and a region,
where $h$ is
dominantly a gauge non-singlet ($c_3$ close to 0), but heavier than
$\sim$ 55 GeV. \par

This feature is also visible in a direct investigation of the
$Z$-$Z$-$h$ coupling, which
is relevant for the process a) of (4). If we denote by $g_h$ the
strength of this coupling
relative to the corresponding coupling in the non-supersymmetric
standard model, we find
that $g_h$ is given by
$$g_h = {c_1<h_1> + c_2<h_2> \over \sqrt{<h_1>^2 + <h_2>^2}} \ \ \ .
\eqno(6)$$

In fig. 3 we plot the logarithm fo $g_h^2$ versus $M_h$ for the present
points in parameter space.
Again we see that for $g_h$ to be close to 1, $M_h$ has to be larger
than $\sim$ 55 GeV, whereas
there exists a long ``tail'' towards lighter Higgs masses, but with very
small coupling $g_h$. As a
dotted line we show in fig. 3 the boundary of the region in this plane,
which has been excluded by
LEP 1 [1] (assuming visible Higgs decays, see below). We see that LEP 1
has actually excluded just a
tiny part of the parameter space. We also show the boundary of the
region which is
visible at LEP 2. Here we define visibility by requiring more than 50
events (before any cuts have
been applied) for a c.m. energy of 175 GeV and an integrated luminosity
of 500 pb$^{-1}$ (dashed line), or for a c.m. energy of 205 GeV and an
integrated luminosity of 300 pb$^{-1}$ (full line).
Of course
the prospects for LEP 2 are better than for LEP 1, but it is also
evident that at least via
the search for a Higgs scalar
even LEP 2 is far from covering the complete parameter space. \par

Let us briefly comment on the decay properties of the lightest Higgs
scalar at this stage. First one
has to check, whether invisible decays into neutralino pairs play a
role. Whereas neutralinos could,
in principle, still be very light within this model [14], we have found
for our range of parameters that within the accessible
region for LEP 2, as in fig. 3, the lightest neutralinos are still
heavier than $M_h/2$, hence this
Higgs decay channel is not open. The lightest Higgs scalar thus decays
practically exclusively
through its $h_1$ component of eq. (5), which couples to $b$ quarks and
$\tau$ leptons with a
relative strength as the standard model Higgs boson. The fact that the
coefficient $c_1$ can be
tiny (for $c_3$ close to 1) decreases, of course, the width of the
lightest Higgs scalar
considerably; its lifetime does not yet become long enough, however,
for allowing it to travel
macroscopic distances such that a displaced vertex could be visible.
Hence the same search
criteria as for the standard model Higgs boson can be applied to the
lightest Higgs scalar for the
present region of parameter space of the (M+1)SSM. \par

Now we turn to the Higgs production process b) of (4). First we
investigate, in fig. 4, which
range of masses of the lightest pseudoscalar boson $A$ as a function of
$M_h$ is allowed. We see
that, within the plotted range, $M_A$ satisfies approximately $M_A \sim
2\cdot M_h$. The allowed
region in the $M_A$-$M_h$ plane is different from the one within the
MSSM [2, 15]. This is not
surprising, however: here the lightest Higgs scalar $h$ is dominantly
gauge singlet, thus in this
region of the parameter space the Higgs sector of the (M+1)SSM
differs substantially from the
MSSM. \par

It turns out, moreover, that for $M_A \leq$ 130 GeV also the lightest
pseudoscalar $A$ is
to more than 99 $\%$ a gauge singlet state. Unfortunately both facts
imply that within the range of
the $M_A$-$M_h$ plane, which is kinematically accessible to LEP~2, the
$Z$-$A$-$h$ coupling is
vanishingly small. The process b) of (4) can thus not be used to test a
part of the present parameter
space. \par

Of course, prospects for the discovery of a light Higgs scalar $h$,
which is dominantly gauge
singlet, look generally quite dim. Fortunately it turns out, however,
that under such circumstances
the second lightest Higgs scalar $H$ cannot be too heavy [5-7]. This
offers some hope to access the
Higgs sector of the (M+1)SSM via this particle. Thus, in fig. 5, we
plot the logarithm of
$g_h^2$ versus the mass $M_H$ of the second lightest Higgs scalar.
Indeed we see that, for small
$g_h$, $M_H$ cannot be too large. The range of $M_H$ corresponding to an
``invisible'' lightest Higgs
scalar $h$, 90 GeV $\leq M_H \leq$ 150 GeV, can, however, hardly be
reached by LEP 2. Below 150 to
160 GeV, on the other hand, a visible Higgs scalar $h$ or $H$ is
guaranteed to exist within this
model, provided the gluino mass (as a measure of the susy breaking
scale) does not exceeed 2 to 3
TeV. \par

Let us return to LEP 2, where the only access to the Higgs sector turned
out to be the process a)
of (4), which can cover the part of the parameter space indicated in
fig. 3. It is of interest to
compare this part of the parameter space with the one, which is
accessible via direct sparticle
searches. In fig. 6 we plot the mass of the lightest chargino versus the
mass of the lightest charged
sleptons (sleptons of different generations are nearly degenerate) for
the range of parameters obtained within our scanning
procedure. We see that, if LEP 2 can detect charginos or charged
sleptons with masses up to $\sim$
90 GeV, an essential part of the parameter space can be tested. We have
to face the question,
whether this part of the parameter space covers completely the one
accessible via the search for a Higgs scalar. \par

In fig. 7 we show the points in the parameter space, which are visible
via the Higgs production
process a) of (4), versus the masses of the lightest chargino and
charged sleptons. Whereas for
most of these points the lightest charginos or the charged sleptons are
indeed lighter than 90 GeV,
we find nevertheless a non-vanishing region in parameter space, in which
the lightest Higgs scalar
can be observed at LEP 2, but both the lightest chargino and the charged
sleptons are too heavy. \par

\noindent 4. Conclusions\par

We can summarize our results as follows. The extension of the MSSM by a
gauge singlet requires a
fresh look at the phenomenology within the Higgs sector. The lightest
neutral Higgs scalar can be
somewhat heavier than in the MSSM. In particular the couplings of the
lightest scalar and
pseudoscalar to the $Z$ boson can be substantially reduced. For the
range of parameters consistent
with universal soft susy breaking terms at $M_{GUT}$ we have found that
a pseudoscalar Higgs is
either too heavy or couples too weakly for LEP 2, and the search for a
Higgs scalar can cover only
a part of the parameter space. If the lightest Higgs scalar is
dominantly gauge singlet and hence
practically invisible, the Higgs boson search has to put up with second
lightest scalar; fortunately, however, this state will then at least be
accessable by the next generation of $e^+$ $e^-$ linear colliders [7].
\par

For the model presented in this paper the search for charginos or
sleptons at LEP~2 seems to be
somewhat more promising than the Higgs boson search; nevertheless a
range of parameters exist, for
which a Higgs boson, but no sparticle would be visible.

 \vfill \supereject
\centerline{\bf \underbar{References}} \par \bigskip
\item{[1]} ALEPH collaboration, Phys. Lett. \underbar{B313} (1993) 312;
\item{} Delphi collaboration, Nucl.
Phys. \underbar{B373} (1992) 3;
\item{} L3 collaboration, Z. Phys. \underbar{C57} (1993) 355;
\item{} Opal
collaboration, Z. Phys. \underbar{C64} (1994) 1;
\item{} J. Rosiek, A. Sopczak, Phys. Lett. \underbar{341} (1995) 419.
\item{[2]} for a review see J. Gunion, H. Haber, G. Kane, S. Dawson, The
Higgs Hunter's Guide (Addison-Wesley, Reading 1990).
\item{[3]} P. Fayet, Nucl. Phys. \underbar{B90} (1975) 104;
\item{} H.-P. Nilles, M. Srednicki, D. Wyler, Phys. Lett.
\underbar{B120} (1983) 346.
\item{[4]} J.-M. Fr\`ere, D. R. T. Jones, S. Raby, Nucl. Phys.
\underbar{B222} (1983) 11;
\item{} J.-P. Derendinger, C. A. Savoy, Nucl. Phys. \underbar{B237}
(1984) 307.
\item{[5]} J. Ellis, J. Gunion, H. Haber, L. Roszkowski, F. Zwirner,
Phys. Rev. \underbar{D39}
(1989) 844;
\item{} M. Drees, Int. J. Mod. Phys. \underbar{A4} (1989) 3635.
\item{[6]} T. Elliot, S. King, P. White, Phys. Lett. \underbar{B305}
(1993) 71, Phys. Rev.
\underbar{D49} (1994) 2435.
\item{[7]} J. Kamoshita, Y. Okada, M. Tanaka, Phys. Lett.
\underbar{B328} (1994) 67.
\item{[8]} U. Ellwanger, M. Rausch de Traubenberg, Z. Phys.
\underbar{C53} (1992) 521.
\item{[9]} U. Ellwanger, M. Rausch de Traubenberg, C. A. Savoy, Phys.
Lett. \underbar{B315} (1993) 331, and in preparation.
\item{[10]} Ph. Brax, U. Ellwanger, C. A. Savoy, preprint DAMTP/94-98,
hep-ph 9411397.
\item{[11]} U. Ellwanger, Phys. Lett. {\underbar{B303} (1993) 271.
\item{[12]} G. Gamberini, G. Ridolfi, F. Zwirner, Nucl. Phys.
\underbar{B331} (1990) 331.
\item{[13]} CDF collaboration, F. Abe et al., Phys. Rev. Lett.
\underbar{73} (1994) 225.
\item{[14]} F. Franke, H. Fraas, A. Bartl, Phys. Lett. \underbar{B336}
(1994) 415.
\item{[15]} J. Ellis, G. Ridolfi, F. Zwirner, Phys. Lett.
\underbar{B262} (1991) 477.
\vfill \supereject \centerline{\bf \underbar{Figure Captions}} \par
\bigskip
{\parindent=1 cm
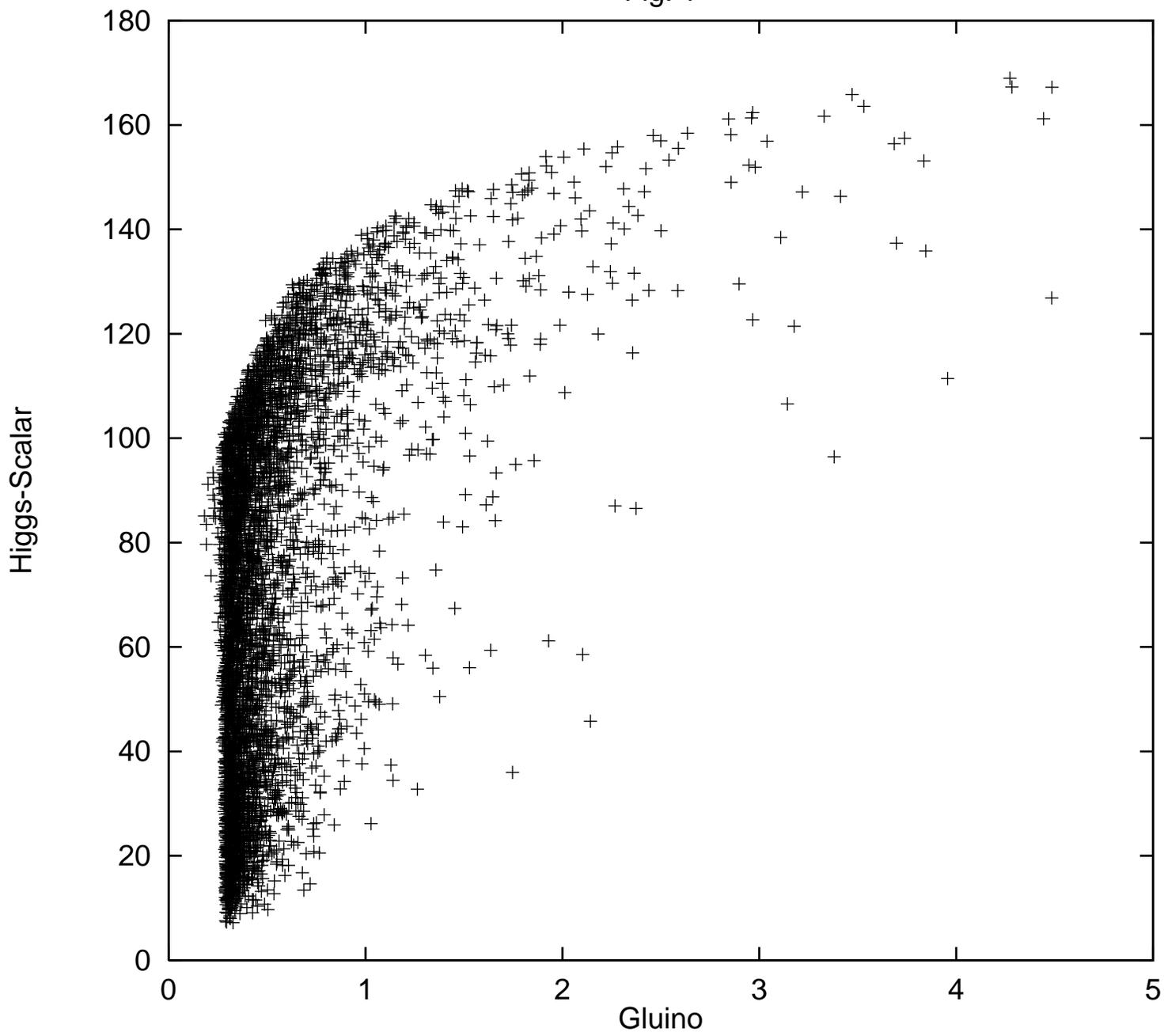
\item{\bf Fig. 1:} Mass of the lightest Higgs scalar (in GeV) versus
the gaugino mass (in TeV).
Unless stated otherwise, the plots are produced using a representative
sample of $\sim$ 5 000
points in the parameter range as described in the text.     \vskip 5 mm
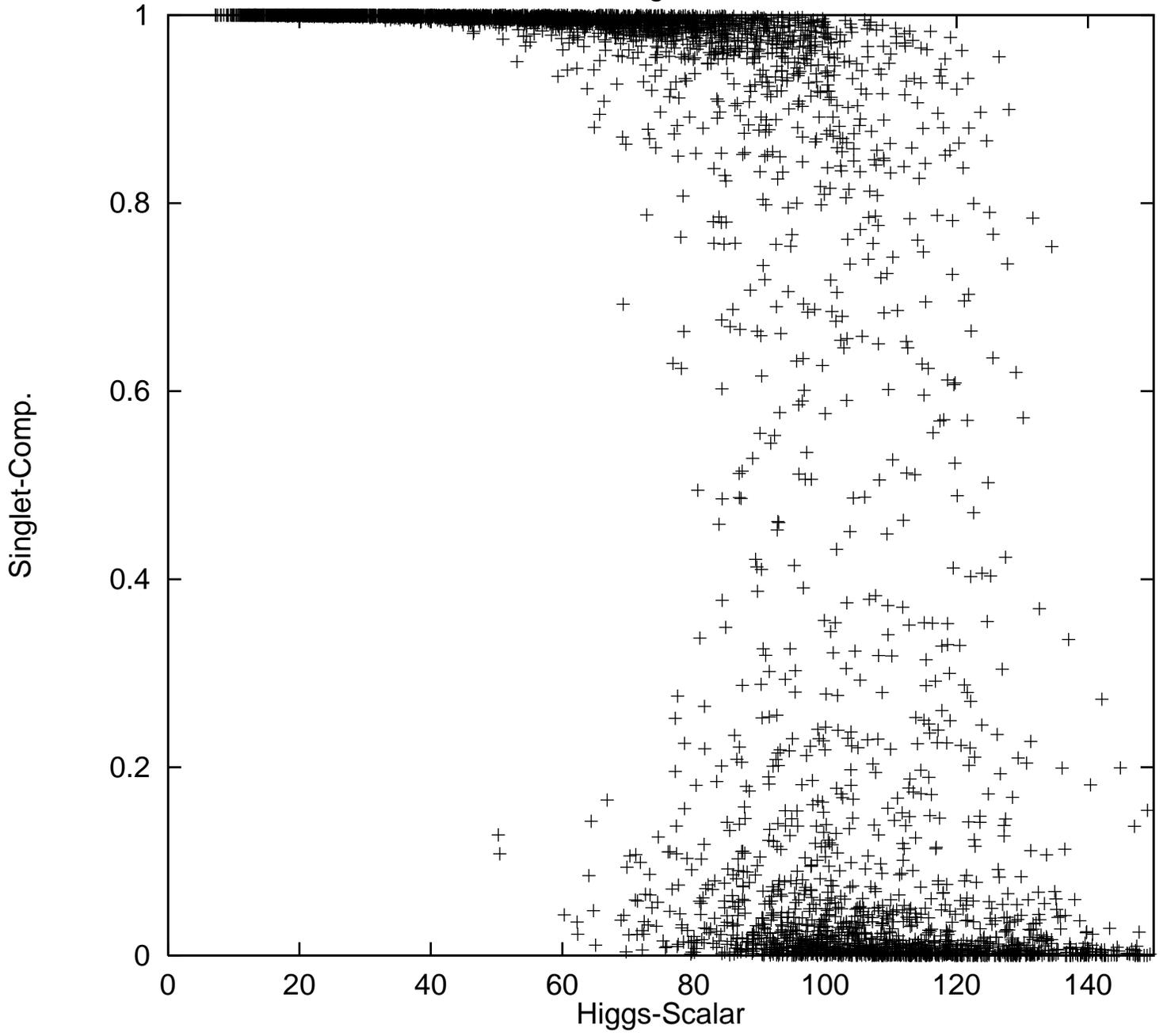
\item{\bf Fig. 2:}
Singlet component $c_3$ of the lightest neutral Higgs scalar versus its
mass (in GeV).       \vskip 5
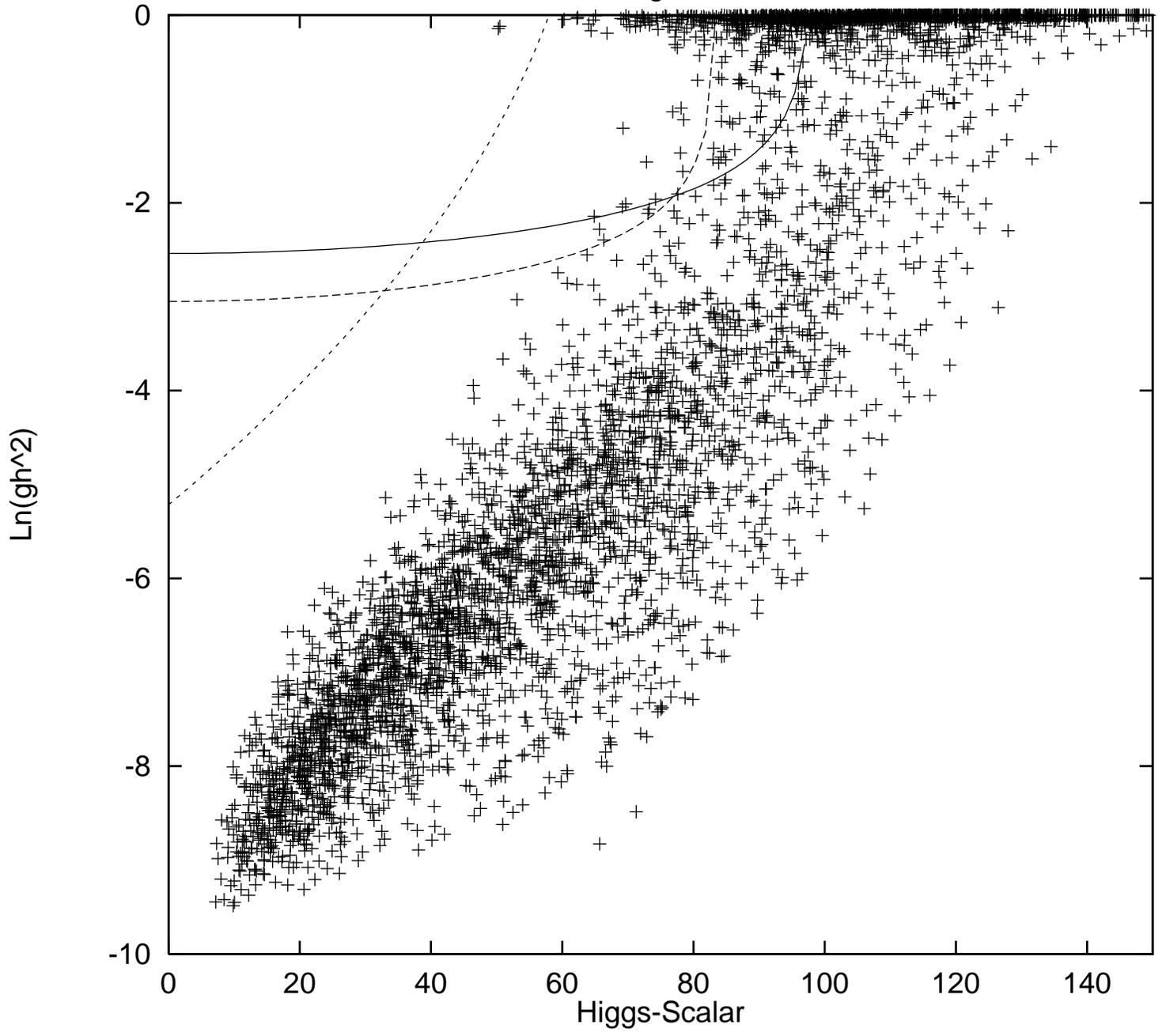
mm \item{\bf Fig. 3:} Logarithm of the $ZZh$ coupling squared versus
$m_h$. Dotted line: region
excluded by LEP 1. Dashed line: region visible at LEP 2
at a c.m. energy of 175 GeV and an integrated luminosity of 500 $pb^{-1}$.
Full line : region visible at LEP 2
at a c.m. energy of 205 GeV and an integrated luminosity of 300 $pb^{-1}$.
\vskip 5 mm
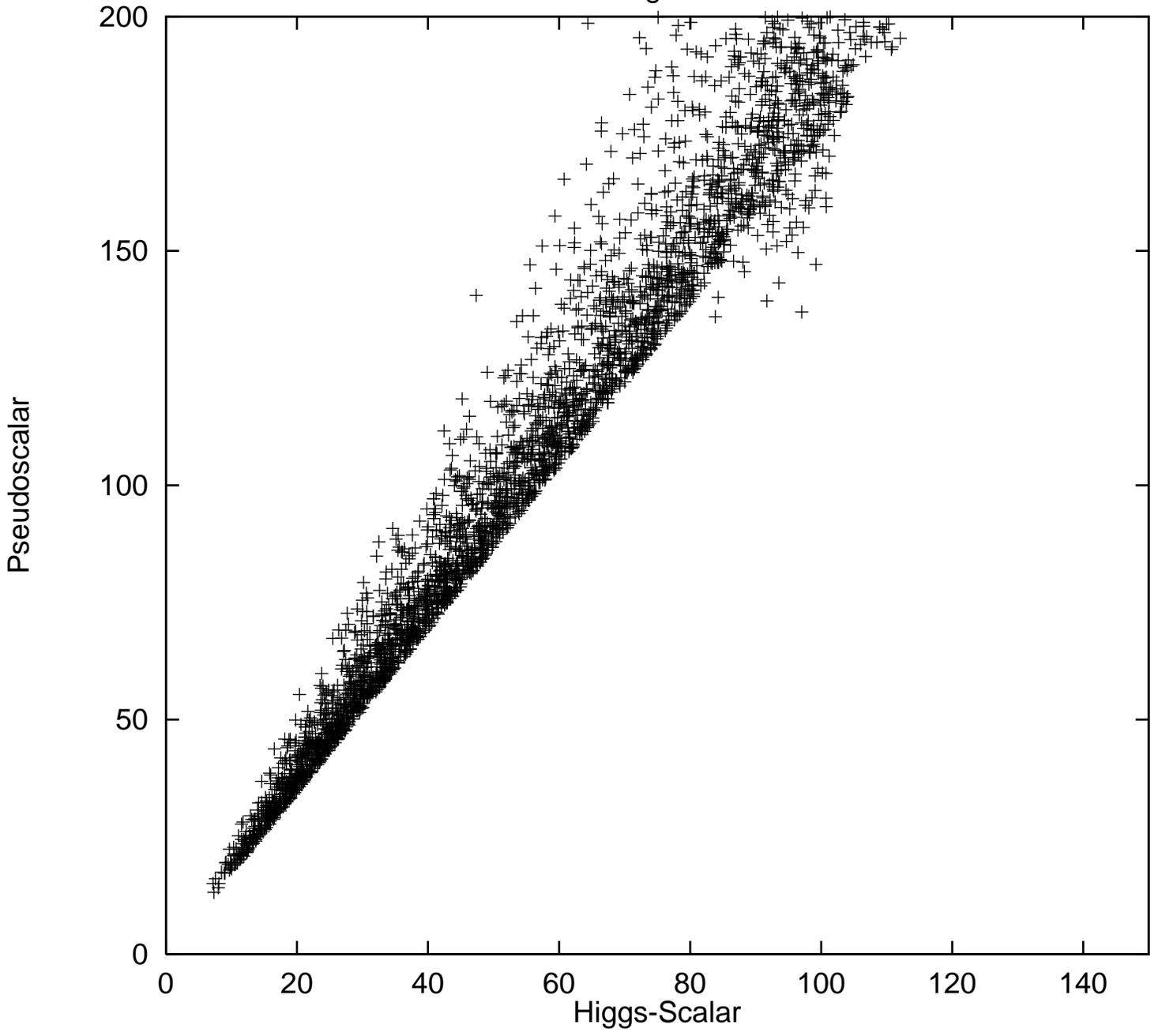
\item{\bf Fig. 4:} Mass of the lightest pseudoscalar $A$ versus $m_h$.
\vskip 5 mm
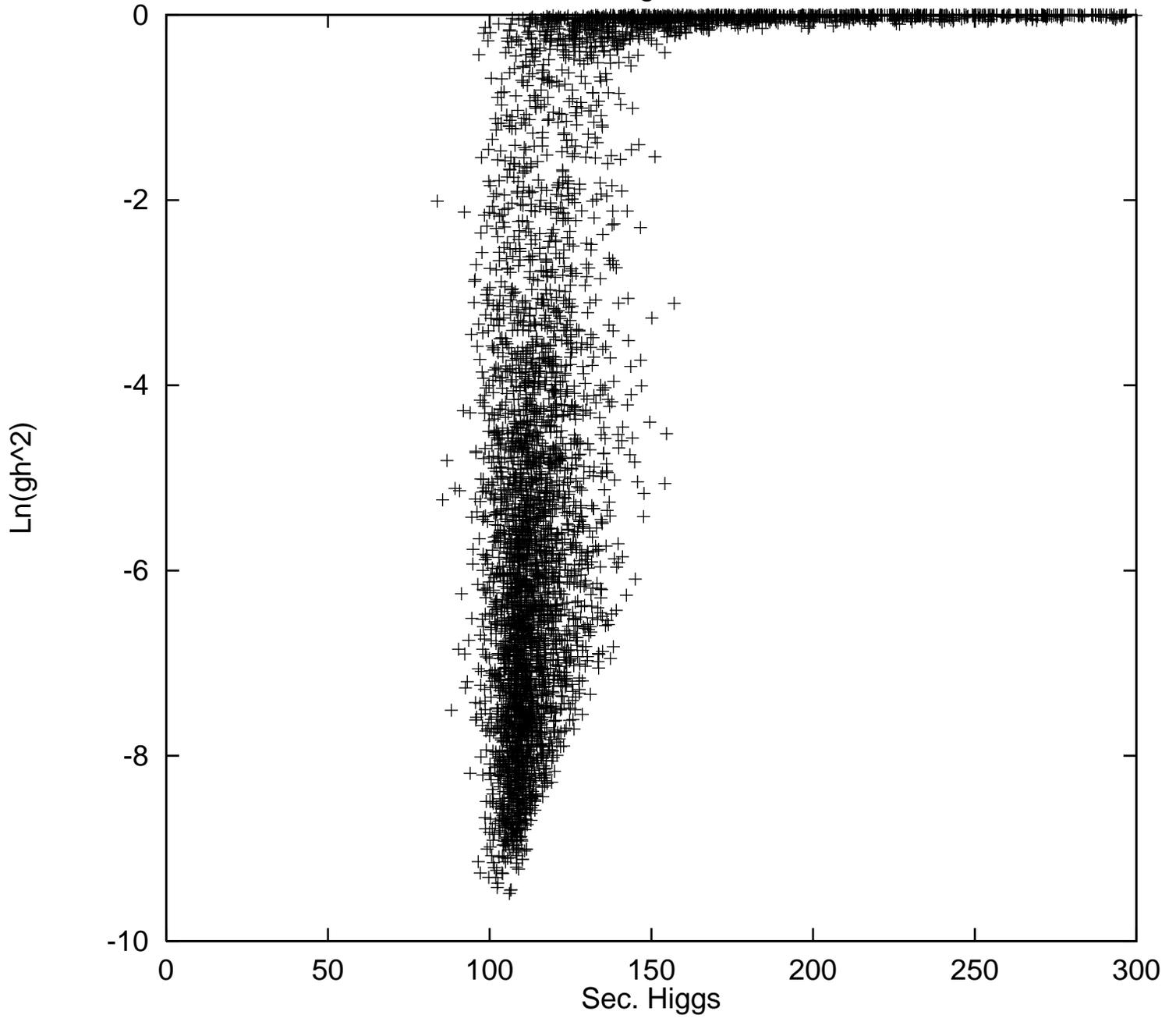
\item{\bf Fig. 5:} Logarithm of the $ZZh$ coupling squared versus the
mass of the second neutral
Higgs scalar.  \vskip 5 mm
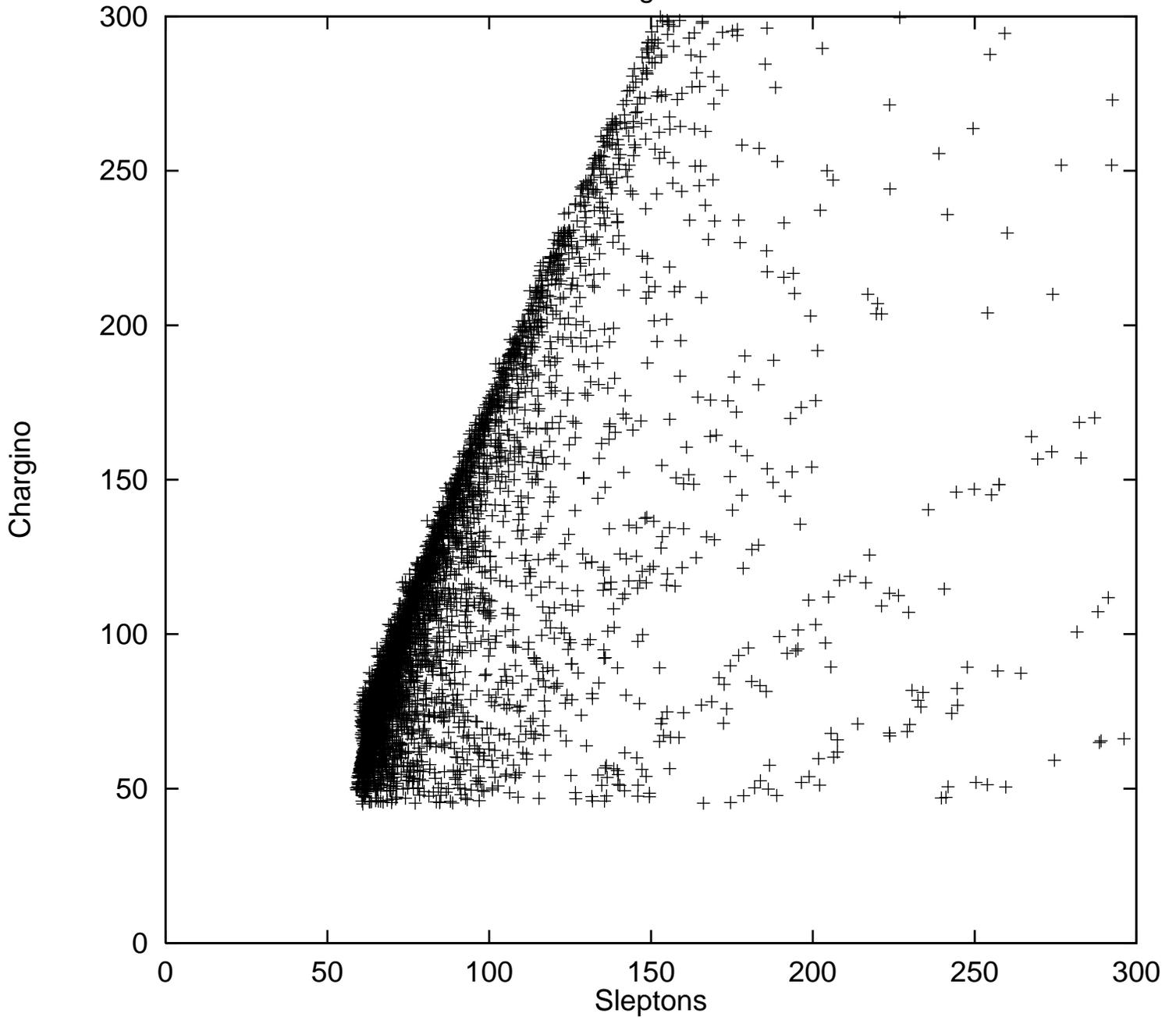
\item{\bf Fig. 6:} Mass of the lightest chargino versus the mass of the
lightest charged slepton.
\vskip 5 mm
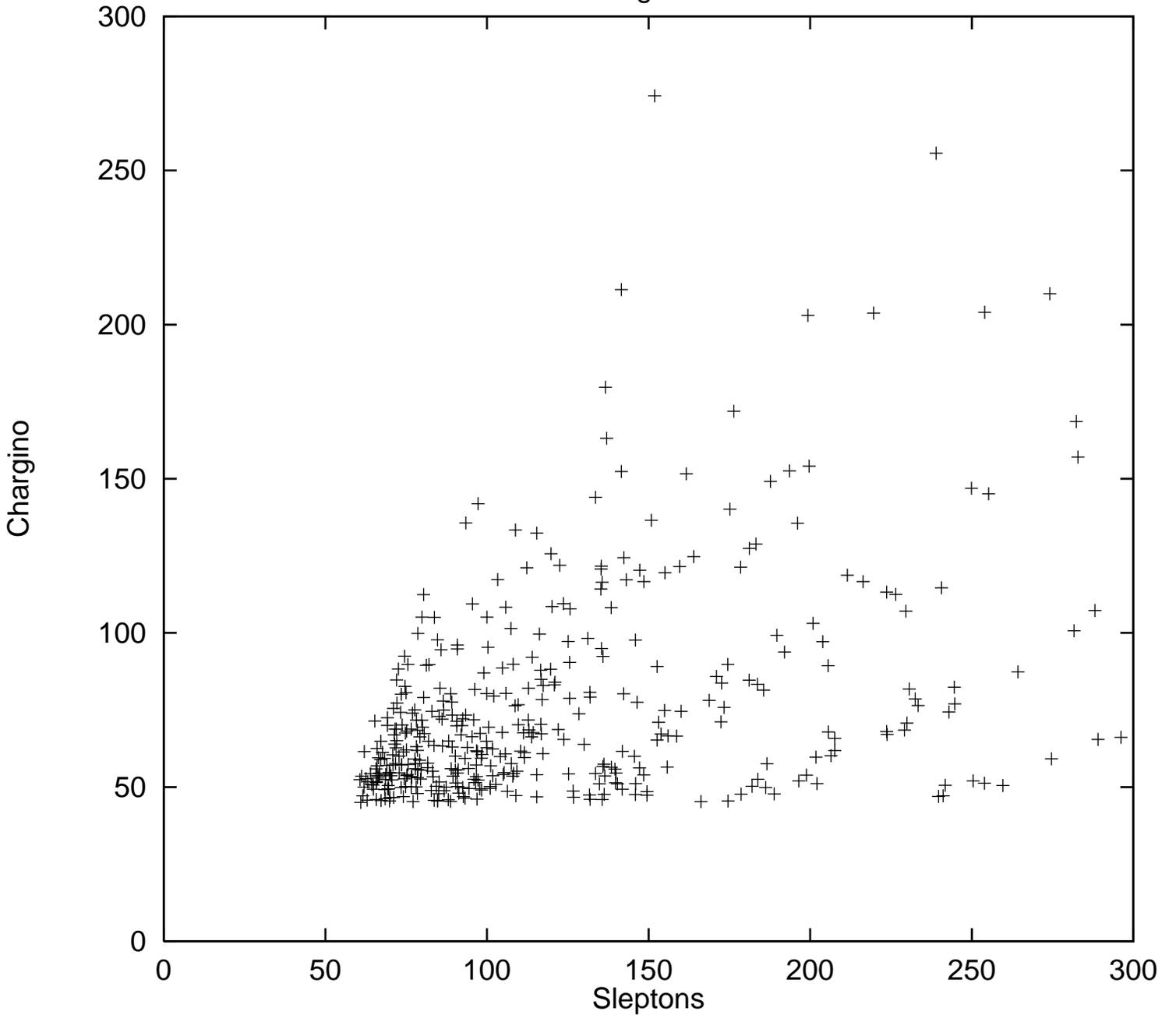
\item{\bf Fig. 7:} As in fig. 6 for only those points in parameter
space, which are visible in
$Z^* \to Zh$ at LEP 2. \par}

 \bye